\documentclass[journal]{IEEEtran}
\ifCLASSINFOpdf
\else
  \usepackage[dvips]{graphicx}
\fi
\begin{document}
%
\title{Effect of dipolar interactions for domain wall dynamics in magnetic thin films}
%
%
%

\author{Adil~Mughal$^{1}$,~
        Lasse~Laurson$^{1}$,~
	Gianfranco~Durin$^{1,2}$,~
        and~Stefano~Zapperi$^{1,3}$\\
	\vspace{0.5cm}
	$^1$ISI Foundation, Viale S. Severo 65, 10133 Torino, Italy.\\
       	$^2$INRIM, Strada delle Cacce 91, 10135 Torino, Italy.\\
	$^3$CNR-INFM, S3, Universit\`a di Modena e Reggio Emilia, Via Campi 
	213/A, 41100 Modena, Italy.}


%
%

\markboth{B1-26}%
{Shell \MakeLowercase{\textit{et al.}}: Bare Demo of IEEEtran.cls for Journals}
%



\maketitle

\begin{abstract}
We study the effect of long range dipolar forces on the dynamics and
morphology of domain walls in magnetic thin films by numerical 
simulations of the spin-1 random field Ising model. By studying
the size distribution of avalanches of domain wall motion arising as a 
response to quasistatic external driving, we observe a cross-over from
the case dominated by short range interactions to another universality
class where the long range dipolar forces become important. This crossover
is accompanied with a change of the domain wall morphology from a
rough wall to walls with zigzag structure. 
\end{abstract}


%
\IEEEpeerreviewmaketitle

\section{Introduction}
%
%
%
%
\IEEEPARstart{T}{he} study of ferromagnetic thin films is an open field of current
interest both from a technological and fundamental point of view. On
the one hand ferromagnetic thin films are of interest because of their
applications in a diverse range of fields, such as magnetic recording
technology and spintronics \cite{Bertotti}, while on the other they are of interest
because there remain a number of unanswered questions regarding their
hysteretic properties, such as the Barkhausen effect.

The Barkhausen effect is the name given to the noise in the magnetic
output generated in a ferromagnetic material when the magnetizing
field applied to it is changed \cite{Barkhausen}. The origin of this noise is due to the
jerky fashion in which magnetic domain walls move in response to a
slowly varying externally applied magnetic field. The motion is
irregular because the domain wall can become pinned, at various
points, by impurities in the material. By increasing the strength of
the external field the domain wall can become locally depinned and
moves forward, only to become trapped once again by more impurities
further ahead \cite{durin05}.

The statistical properties of the Barkhausen effect are quantitatively
understood in bulk three dimensional materials by the theory of domain
wall depinning \cite{Urbach,Zapperi1}. The experiments can be
classified into universality classes depending on the strength of
dipolar interaction \cite{Zapperi2}. On the other hand, our
understanding of Barkhausen noise in ferromagnetic thin films is
incomplete. Recent experiments show that the morphology of the domain
wall depends strongly on temperature \cite{Ryu}. At temperatures well
below the Curie temperature, with high saturation magnetization, the
domain wall is observed to have a zigzag structure to minimise the the
dipolar interaction energy. However, upon approaching the Curie
temperature the morphology of the domain wall is dominated by line
tension and is observed to form a rough interface free of large
zigzags. These two regimes correspond to different Barkhausen noise
universality classes and the exponents exhibit a crossover as a
function of temperature.

Here we demonstrate that this crossover can be understood by
considering the effect of the long range dipolar forces on the
morphology and dynamics of a magnetic domain wall in a thin
ferromagnetic material. We consider a head-on domain wall consisting
of two domains of opposite in plane magnetisation, due to large
anisotropy along the easy axis. This situation we model as a three
state spin-1 Ising Hamiltonian \cite{Iglesias}, whereby the interaction
between neighbouring spins is ferromagnetic and is in competition
with a long range (antiferromagnetic) dipole-dipole interaction. In
addition the Hamiltonian also includes a random quenched field (which
represents the effect of pinning sites) and an extra term which models
the effect of an external magnetic field. By increasing the magnitude
of the external field we can drive the domain wall forward over the
pinning sites and collect statistics for the avalanche size
distributions. Doing this for a range of values of the saturation
magnetisation we confirm that behaviour of the Barkhausen noise can be
described by two separate universality classes, depending on the value
of the saturation magnetization. The paper is organized as follows: In
the next Section we present the simulation model, followed by an
overview of the numerical results in Section III. Section IV finishes
the paper with conclusions.


 

\section{Model}

We simulate the dipolar spin 1 random field Ising model (RFIM) in a 2D triangular
lattice  with in plane magnetization. A similar model was studied in 
Ref.~\cite{Iglesias} with out of plane magnetization. The Hamiltonian is given by
\begin{eqnarray}
\label{eq:H}
\mathcal{H} & = &-J\sum_{\langle ij \rangle} s_i s_j -
\sum_i [(H+h_i)s_i-\frac{1}{2}As_i^2] +\\ \nonumber 
& & +D \sum_{ij}s_i s_j\frac{1-3\cos^2 \theta_{ij}}{r_{ij}^3},
\end{eqnarray}

where the first term describes the exchange interactions between
nearest neighbour spins with a strength $J$, the second term takes
into account the contributions of the applied external field $H$ and
the parameter $A$ controls the relevance of the $s_i=0$ state,
respectively.  The third term includes the dipolar interactions
between two spins $s_i$ and $s_j$ at a distance $r_{ij}$ , with a
strength $D=\mu_0 M_s^2/2\pi$, where $M_s$ is the saturation
magnetization.  In the simulations, we also add a fictitious demagnetizing
energy $E=1/2  k M^2$, with $k$ the demagnetizing factor and
$M=M_s\sum_i s_i/N$ the total magnetization of the system. This final
term has been introduced only for computational convenience, since dipolar
interactions are accounted for by the third term, and allows the
domain wall to be kept around the coercive field. In experiments
this situation is realised by recording magnetization dynamics slightly below the
coercive field $H_c$ and then letting the domain wall move by thermal
activation. This case would be slower to simulate with our model, but avalanche
statistics is expected to be same in the limit of small $k$. Finally, we note
that disorder is modeled by in the system by a quenched random field
$h_i$ extracted from a Gaussian distribution with variance $R$.

\begin{figure}[!t]
\centering
\includegraphics[width=2.85in]{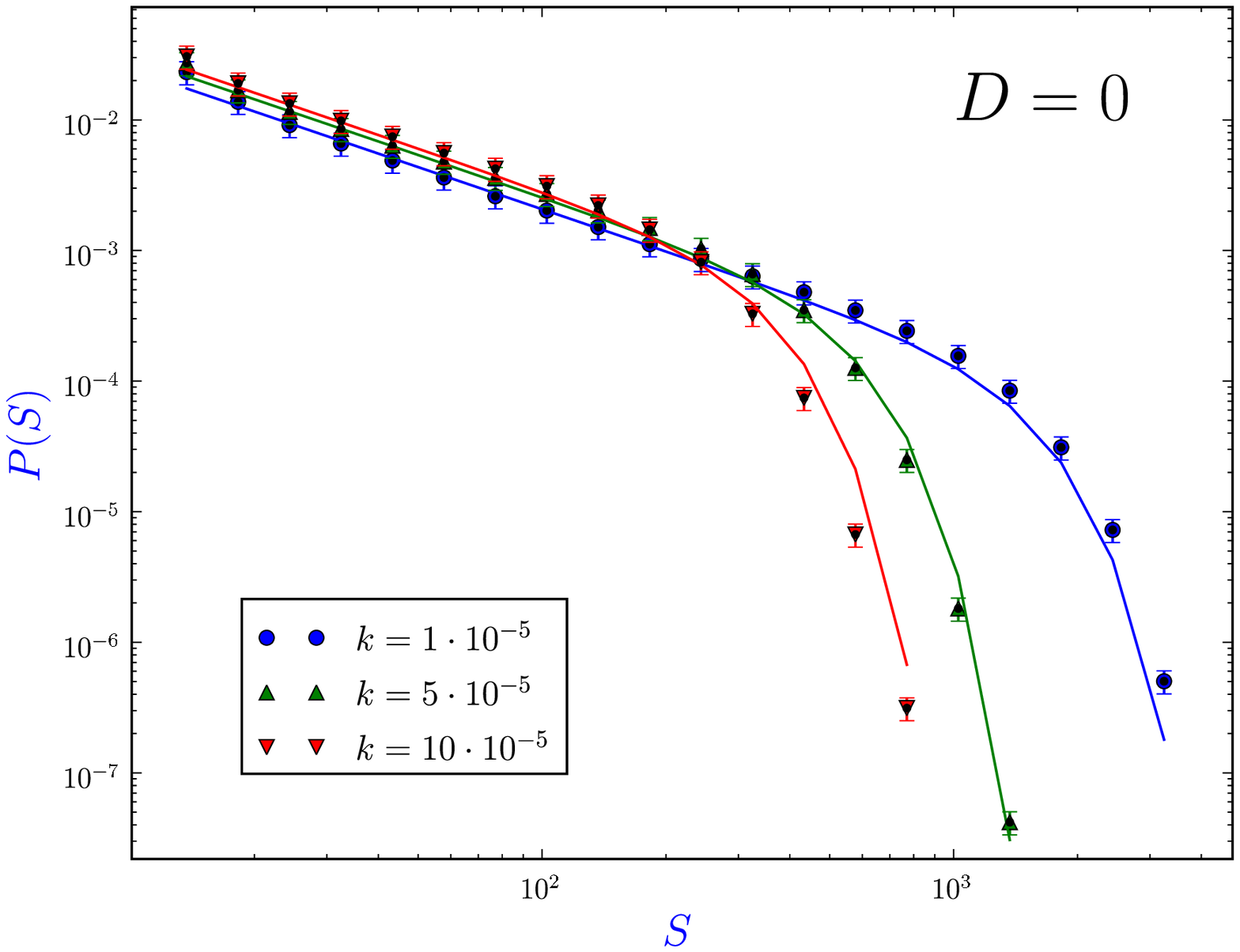} \\
\includegraphics[width=2.85in]{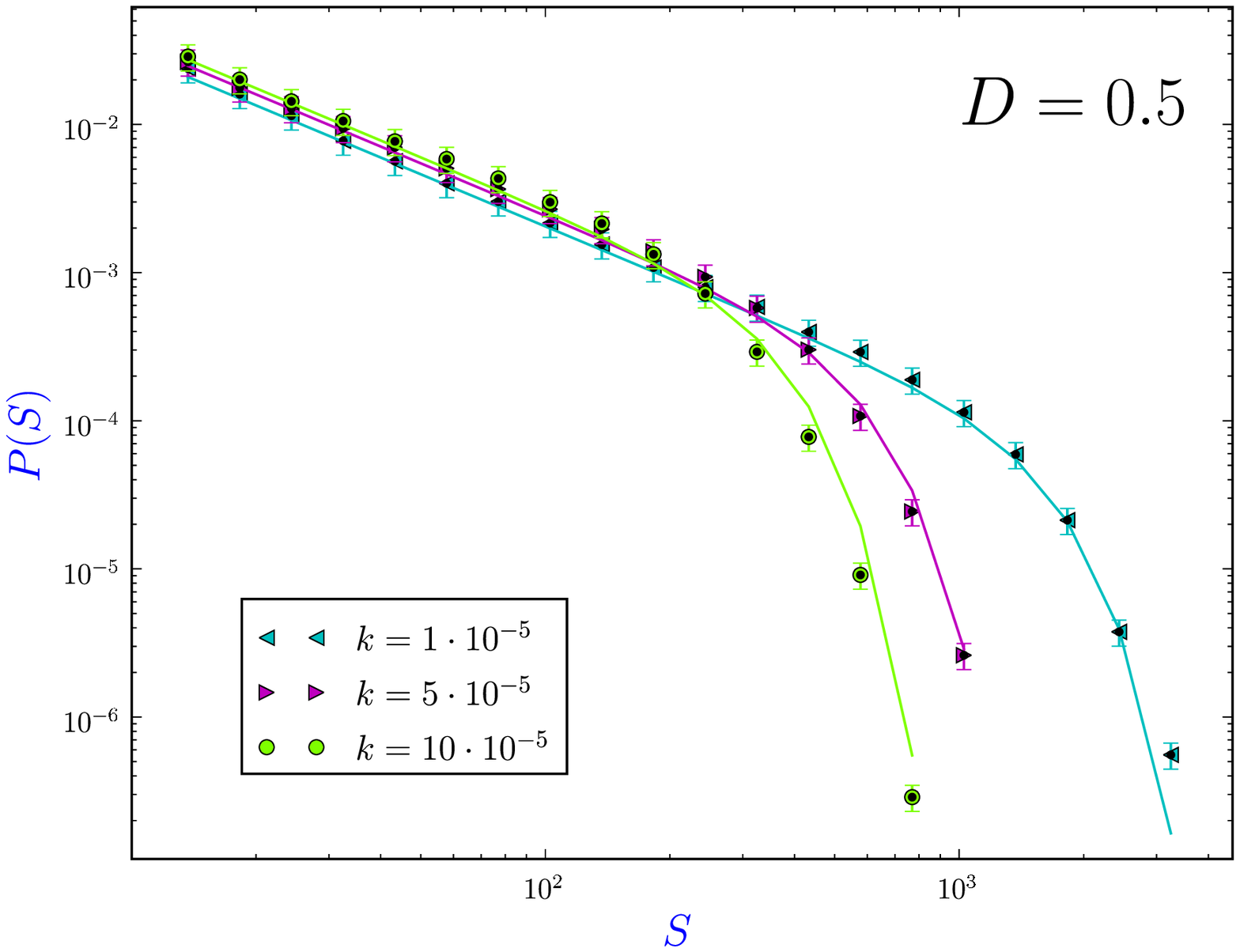} \\
\includegraphics[width=2.85in]{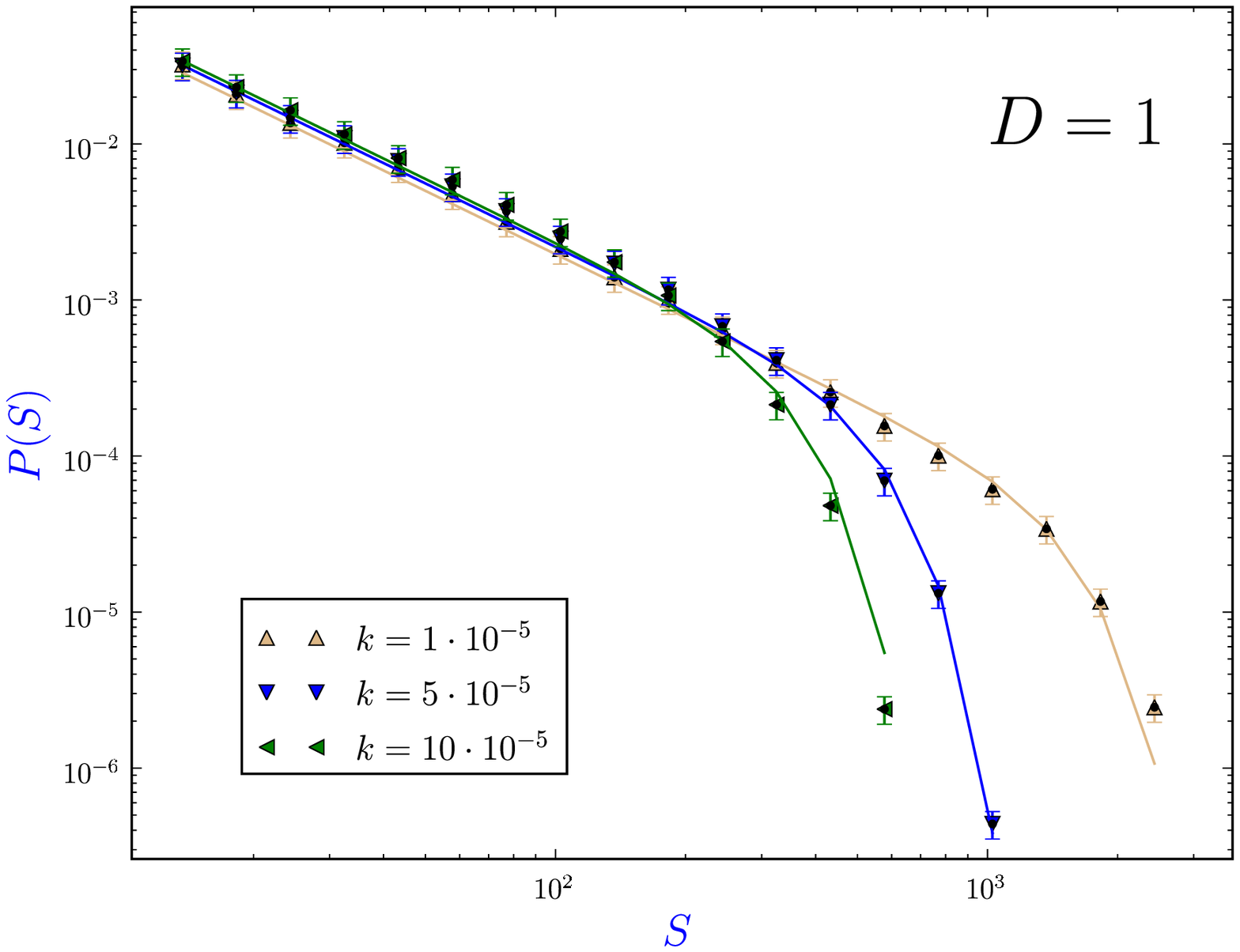} \\
                                                                                           
                                                                                           
                                                                                           
\caption{The avalanche size distributions for different strengths of the
dipolar interaction, along with fits of the form of Eq. (\ref{eq:pl}).
The top panel shows the case with $D=0$ (with $\tau=1.07\pm0.03$,
$\sigma_k=0.64\pm 0.02$ and $n=3.1\pm 0.4$), followed by
$D=0.5$ (with $\tau=1.17\pm0.03$,
$\sigma_k=0.64\pm 0.02$ and $n=2.6\pm 0.3$) and $D=1$ (with $\tau=1.35           
\pm0.03$, $\sigma_k=0.64\pm 0.02$ and $n=2.5\pm 0.2$). 
}
\label{fig:fits}
\end{figure}

To study the avalanche dynamics and the morphology of the domain
walls, we simulate the model on a two dimensional lattice of size $L
\times 2L$, with $L=256$, in $T=0$ with extremal dynamics (meaning
that after each avalanche the external field is ramped up to the
point where just a single spin is about to flip, the spin is then
allowed to flip and this initiates the next avalanche). We consider
weak disorder, setting $R/J=1/2$. 

In this limit, a spin-1 model is convenient in describing the domain
wall dynamics since the intermediate $s_i=0$ state causes the domain
walls to have a finite width so that spurious lattice effects are less
pronounced than in systems where an abrupt transition between
neighboring domains takes place. Such lattice effects, which are a
problem in the spin $\frac{1}{2}$ case, include faceting a case in
which the length of a boundary between two points has a different
value depending on its orientation with respect to the
crystallographic axis \cite{ji91}. Lattice anisotropy is also reduced
using a triangular lattice, instead of the more conventional square
lattice. Finally, the parameter $A$ is chosen so that problem of
faceting, as described above, is minimized - which according to our
simulations occurs for $A \simeq 2J$.  The simulations are started
from a state with all spins $s_i=-1$ except at the bottom layer of the
system where a boundary condition $s_i=1$ is imposed. This creates an
initially flat head-to-head domain wall in the system. The external
field value is then increased from zero until the first spin at the
domain wall becomes unstable, i.e. the local field $H_i$ of the spin
$s_i$ given by
\begin{eqnarray}
H_i & = & J\sum_{\langle ij \rangle}s_j + H + h_i -A s_i- \\ \nonumber 
& & -D \sum_{ij}s_j\frac{1-3\cos^2 \theta_{ij}}{r_{ij}^3} - kM
\end{eqnarray} 
changes sign (note here we have explicitly included the fictitious
demagnetizing term). This spin is then flipped from $s_i=-1$ to
$s_i=0$, or from $s_i=0$ to $s_i=1$. Due to the interactions between
different spins (both through the local exchange interactions and the
long range dipolar fields), this initial spin flip can cause other
spins to flip as well, leading to a cascade of activity or an
avalanche. The external field is kept constant during such an
avalanche.  The avalanche size $S$ is defined as the total number of
spins flipped during an avalanche. All the spin flips are taken to
proceed via the intermediate $s_i=0$ state, and only spins at the
domain wall are flipped, preventing domain nucleation in front of the
interface. Each time the activity stops, the magnitude of the external
field $H$ is again increased by an amount which causes one of the
spins on the lattice to become unstable, and a new avalanche is
initiated. This procedure is repeated until the domain wall has moved
from one end of the system to the other. Notice that the inclusion of
the term describing the effects of the demagnetizing fields keeps the
domain wall close to the critical point of the underlying depinning
transition as the external field $H$ is ramped up.

\begin{figure}[!t]
\centering
\includegraphics[width=1in]{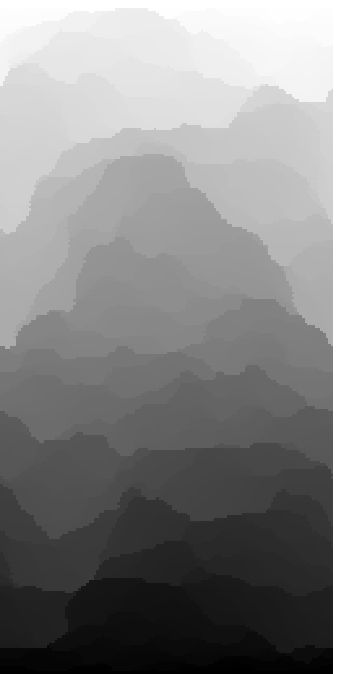}
\hspace{0.05in}
\includegraphics[width=1in]{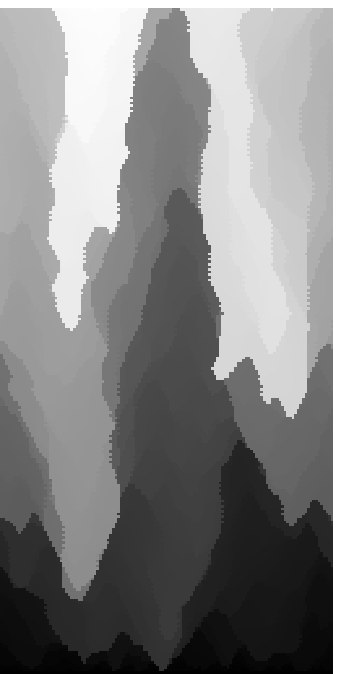}
\hspace{0.05in}
\includegraphics[width=1in]{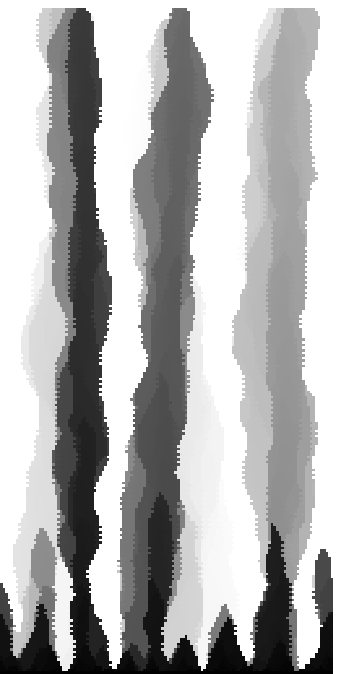}
\caption{Examples of domain evolution patterns for $D=0$ (left),
$D=0.5$ (middle) and $D=1.0$ (right). The different
grayscale colors correspond to time (flowing from black to white) such that
the area of each avalanche has its own grayscale. Notice the change in the
domain wall morphology as the strength of the dipolar forces is increased.}
\label{fig:domain}
\end{figure}

\section{Numerical results}

Simulations are performed with $R=1$, $J=2$, $A=4$, three different values of the 
dipolar interaction strength ($D=0, 0.5$ and $1$) and of the demagnetizing
factor $k$. Notice that $k$ and $D$ should in principle be related. Here, we vary
them independently for computational purposes, since $k$ alone is expected
to determine the avalanche cut-off \cite{Zapperi2}.
The avalanche size distributions are fitted by a power law with a cut-off,
\begin{equation}
\label{eq:pl}
P(S) = C S^{-\tau} \exp{\left[-\left(\frac{S}{S_0}\right)^n\right]},
\end{equation}
where $C$ is a normalization constant, $\tau$ is the exponent characterizing the universality class of
the avalanche dynamics, $S_0 = k^{-\sigma_k}$ gives the dependence of the
cut-off scale $S_0$ on the demagnetizing factor $k$ \cite{Zapperi2}, and $n$ is a fitting
parameter related to the shape of the cut-off function. Fig. \ref{fig:fits} shows
the distributions for the three values of $D=0, 0.5$ and $1$ and three values of $k$.
For each value of $D$, we perform a simultaneous least-square minimization on the three curves 
corresponding to the three values of $k$.
The best least-squares fits to the data give $\tau=1.07 \pm 0.03$ for $D=0$,
$\tau=1.17 \pm 0.03$ for $D=0.5$, and $\tau=1.35 \pm 0.03$ for $D=1$.
The other parameter values are $\sigma_k=0.64\pm 0.02$ for all three cases, and
$n=3.1 \pm 0.4$, $n=2.6\pm 0.3$ and $n=2.5\pm 0.2$ for $D=0$, $0.5$ and $1$, 
respectively.  The values of the avalanche size exponent $\tau$ are in excellent agreement with
the results reported in Ref.~\cite{Ryu} for MnAs thin films, reporting a crossover
between $\tau=1.33$ and $\tau \simeq 1$ as the saturation magnetization $M_s$ was decreasing
due to the temperature increase. 
In addition, the exponent for the case without  dipolar interactions is close to the
theoretically expected value of $\tau \simeq 1$ for the $1+1$ dimensional interface
depinning of the short range interaction universality class \cite{Fisher,Narayan}. 
When the strength 
of the dipolar interactions is increased, we observe a crossover to a different 
exponent value, close to the value reported for a lattice model
for zigzag domain walls \cite{Cerruti}.

By inspecting the domain wall morphology in Fig. \ref{fig:domain} we observe 
that this change in the exponent value is accompanied with a change in 
the domain wall structure: For
low $D$, the domain walls are almost flat, with some roughness due
to the interaction with the random field disorder. For larger values of
$D$, a characteristic zigzag or sawtooth pattern develops, as
a result of the interface trying to minimize the magnetostatic energy due
to the magnetic charges at the interface. This feature is again in excellent 
agreement with the experiment \cite{Ryu}.

\section{Conclusions}

Our simulations demonstrate a striking change in the morphology of a
head-on domain wall as the saturation magnetisation is increased. It
is clear that this change in the morphology is responsible for the
cross-over between the two universality classes, as evidenced by the
change in value of the avalanche size exponent $\tau$. This difference helps
explain, in part, the variation in the experimentally measured
scaling exponents.

However, despite this success we believe that the present model is
subject to limitations due to the underlying lattice symmetry of the
model. It is well known that lattice effects can have an unwanted
influence on the dynamics/morphology of a system and this can only be
overcome by simulating prohibitively large systems. In order to obtain
a faithful estimation of the avalanche size exponents it is necessary
instead to consider an 'off-lattice' model. In such a model the object
of interest is the domain wall itself, which can be described as a
series of vertices connected by straight edges. The various
interactions between spins can be interpreted as forces acting on the
domain wall, while the position of the vertices describing the domain
wall become the degrees of freedom of the system. Such a model is
currently under investigation by the present authors.

\section*{Acknowledgment}

LL wishes to thank Academy of Finland for financial support.

\ifCLASSOPTIONcaptionsoff
  \newpage
\fi

\end{document}